\documentclass[sigconf,nonacm]{acmart}
\usepackage{soul}

\usepackage{listings}
\usepackage{xcolor}

\colorlet{punct}{red!60!black}
\definecolor{background}{HTML}{EEEEEE}
\definecolor{delim}{RGB}{20,105,176}
\colorlet{numb}{magenta!60!black}

\lstdefinelanguage{json}{
    basicstyle=\normalfont\ttfamily,
    numbers=left,
    numberstyle=\scriptsize,
    stepnumber=1,
    numbersep=8pt,
    showstringspaces=false,
    breaklines=true,
    frame=lines,
    backgroundcolor=\color{background},
    literate=
     *{0}{{{\color{numb}0}}}{1}
      {1}{{{\color{numb}1}}}{1}
      {2}{{{\color{numb}2}}}{1}
      {3}{{{\color{numb}3}}}{1}
      {4}{{{\color{numb}4}}}{1}
      {5}{{{\color{numb}5}}}{1}
      {6}{{{\color{numb}6}}}{1}
      {7}{{{\color{numb}7}}}{1}
      {8}{{{\color{numb}8}}}{1}
      {9}{{{\color{numb}9}}}{1}
      {:}{{{\color{punct}{:}}}}{1}
      {,}{{{\color{punct}{,}}}}{1}
      {\{}{{{\color{delim}{\{}}}}{1}
      {\}}{{{\color{delim}{\}}}}}{1}
      {[}{{{\color{delim}{[}}}}{1}
      {]}{{{\color{delim}{]}}}}{1},
}

\AtBeginDocument{%
  \providecommand\BibTeX{{%
    \normalfont B\kern-0.5em{\scshape i\kern-0.25em b}\kern-0.8em\TeX}}}

\setcopyright{acmcopyright}
\copyrightyear{2022}
\acmYear{2022}
\acmDOI{}

\acmConference[]{}{}{}

\usepackage{xspace}
\newcommand{\AlgName}{{CTI4AI}\xspace}

\acmPrice{}
\acmISBN{}

\begin{document}

\title{\AlgName: Threat Intelligence Generation and Sharing after Red Teaming AI Models}


\author{Chuyen Nguyen}
\affiliation{%
  \institution{Mississippi State University}
  \country{Mississippi State, MS, USA}
  }
\email{cjn146@msstate.edu}

\author{Caleb Morgan}
\affiliation{%
  \institution{Mississippi State University}
  \country{Mississippi State, MS, USA}
  }
\email{ctm365@msstate.edu}

\author{Sudip Mittal}
\affiliation{%
  \institution{Mississippi State University}
  \country{Mississippi State, MS, USA}
}
\email{mittal@cse.msstate.edu}

\renewcommand{\shortauthors}{Nguyen and Morgan et al.}

\begin{abstract}

As the practicality of Artificial Intelligence (AI) and Machine Learning (ML) based techniques grow, there is an ever increasing threat of adversarial attacks. There is a need to `red team' this ecosystem to identify system vulnerabilities, potential threats, characterize properties that will enhance system robustness, and encourage the creation of effective defenses. A secondary need is to \textit{share} this AI security threat intelligence between different stakeholders like, model developers, users, and AI/ML security professionals. In this paper, we create and describe a prototype system \AlgName, to overcome the need to methodically identify and share AI/ML specific vulnerabilities and threat intelligence.
\end{abstract}



 \maketitle

\section{Introduction \& Background}

As the practicality of Artificial Intelligence (AI) and Machine Learning (ML) based techniques grow, there is an ever increasing threat of adversarial attacks. This universal AI security issue has prompted a need for methods and measures to `red team' this ecosystem. In 2019, DARPA first recognized this need for an AI/ML centric defense testing toolkit and introduced the Guaranteeing AI Robustness Against Deception (GARD) framework \cite{darpaFound}. GARD aims to identify system vulnerabilities, characterize properties that will enhance system robustness, and encourage the creation of effective defenses.
GARD uses a variety of toolboxes such as Armory, APRICOT (Adversarial Patches Rearranged In COnText), and Adversarial Robustness Toolbox (ART) \cite{darpaFound,gardtoolkits}. Of these toolkits - 
ART, provides a variety of adversarial attacks for a developer to red team their machine learning models. ART allows adversarial attacks such as evasion, poisoning, extraction, and inference \cite{trusted-ai}. It also allows a developer to understand defenses, estimators, and metrics to measure the robustness and effect of these attacks against the model.

A secondary need is to \textit{share} this AI security information between different stakeholders like, model developers, users, and AI/ML security professionals. The need for sharing can arise at the time of model training, testing (security or otherwise), and also during deployment when previously unknown vulnerabilities are found, mitigated or patched. Several existing popular Cyber Threat Intelligence (CTI) sharing techniques and platforms like, OASIS Trusted Automated Exchange of Intelligence Information (TAXII) \cite{taxii} (based on Structured Threat Information Expression (STIX)) \cite{stix1}, MISP Threat Sharing ecosystem \cite{misp}, and traditional sources like National Vulnerability Database \cite{nvd}, etc. allow cybersecuity professionals to share and disseminate software and hardware vulnerabilities and threats. A similar pipeline is needed for the dissemination of AI/ML model specific vulnerabilities. The first step required to build such a pipeline is the creation of a knowledge base containing observations from AI/ML red teams like, AI adversary Tactics, Techniques, and Procedures (TTPs). One such early effort is {MITRE ATLAS (Adversarial Threat Landscape for Artificial-Intelligence Systems) knowledge base} \cite{mitreatlas} modeled after the MITRE ATT\&CK framework \cite{mitreattack}. ATLAS includes a well-defined overview of adversary tactics, techniques, and case studies for AI systems based on real-world observations and demonstrations from AI security groups, and from academic research.

\begin{figure*}[h]
    \centering
    \includegraphics[scale=0.35]{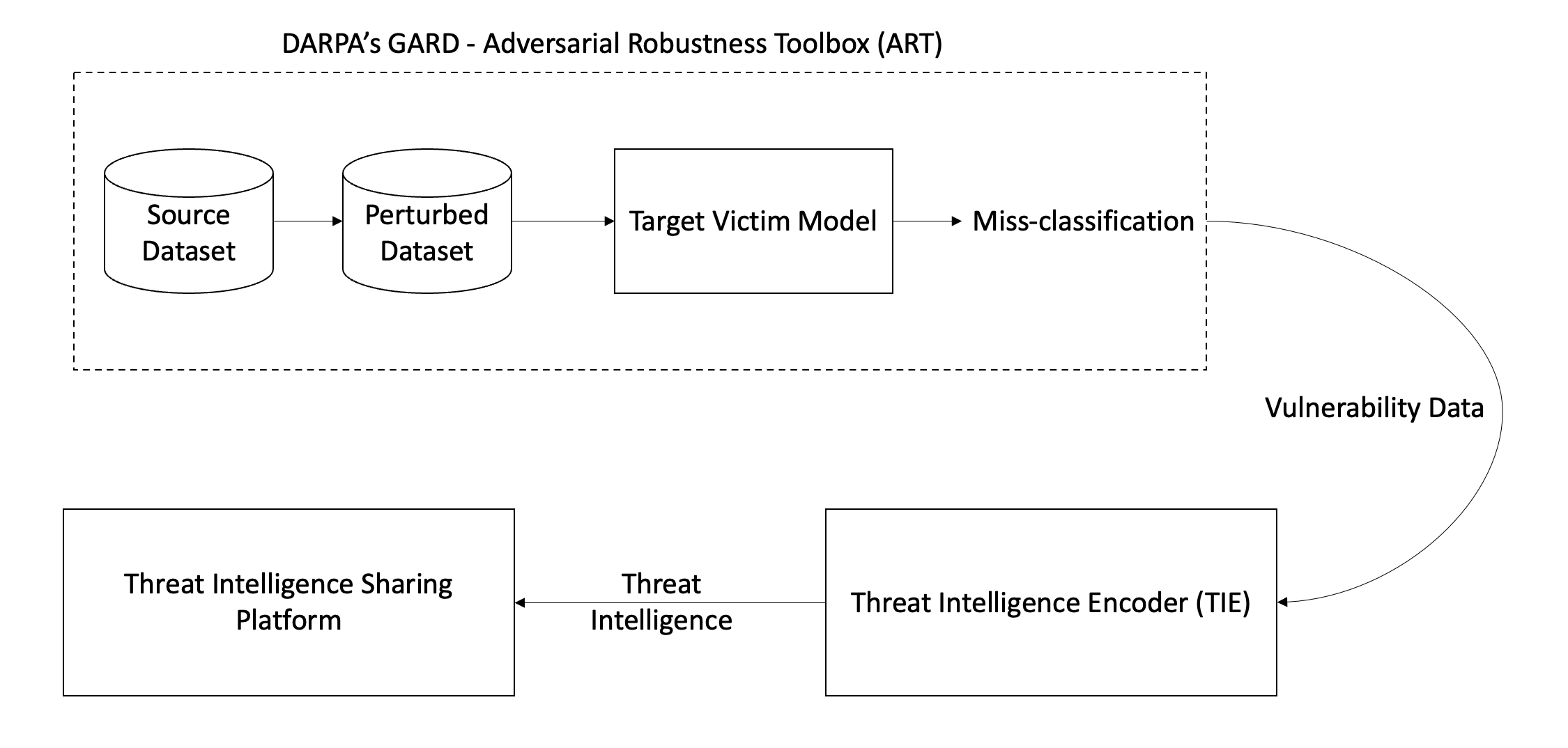}
    \caption{CTI4AI Architecture detailing the three system sub-components.}
    \label{fig:arch}
\end{figure*}

In this paper, to overcome the need to methodically identify and share AI/ML specific vulnerabilities and threat intelligence we create and describe a prototype system \AlgName. The system leverages DARPA's GARD AI red teaming toolkit to identify vulnerabilities in an AI model. These vulnerabilities and threat intelligence are encoded in an adapted STIX object, called Artificial Intelligence Threat Information (AITI), that can then be shared using a RESTful API server like TAXII \cite{taxii}. The architecture of the \AlgName prototype system has been described in Section \ref{system} (also see Figure \ref{fig:arch}).

To showcase the overall system, we also present a case study where the \AlgName system is used to identify and share a vulnerability in an object identification AI model, built using the Resnet-50 architecture \cite{he2016deep}, trained on the CIFAR-10 dataset \cite{krizhevsky2009learning}. 

We use the Fast Gradient Method (FGM) tool part of the ART toolkit. FGM is a white-box adversarial method first introduced in 2015 \cite{goodfellow_shlens_szegedy_2015}. The main idea of FGM is to apply a small perturbation to an input image to cause a machine learning model to misclassify. This is done in three steps: 1) Pass the image through the network, 2) Calculate the loss, 3) Apply a gradient to the pixels of the image in order to \textit{maximize} the loss. Targeted FGM refers to the usage of FGM to trigger a specific misclassification, whereas untargeted FGM only attempts to cause the model to misclassify. It is important to note that the general goal of FGM is to cause a loss that is opposite of the network's goal. This means that if the goal for the network to minimize loss, then the adversarial method should maximize loss, and vice versa.
    
The target victim model used as example for attack in this study is the Resnet-50 model, a residual convolutional neural network popular in usage for simple image-based tasks \cite{he2016deep}. Resnet-50 was chosen for its popularity and well-documented performance, allowing for it to serve as a baseline for demonstrating our \AlgName system and architecture. The CIFAR-10 dataset was used to train and evaluate the Resnet model as well as generate adversarial images to attack it \cite{krizhevsky2009learning}. The particular dataset was chosen as it is a common benchmark for many image-based models, and for the fact that ART also implements a loader class to utilize it in its tool-set.
    
\section{CTI4AI Architecture}\label{system}
The \AlgName system architecture, including three sub-components has been showcased in Figure \ref{fig:arch}. First, we generate AI model vulnerability data by red teaming a target victim model.  
Next, we will take the vulnerability data produced and pass it to the second subsystem called Threat Intelligence Encoder (TIE). This encoder can create a standard representation for this data in the Artificial Intelligence Threat Information (AITI) format. Finally, we can take the output from the Threat Intelligence Encoder, and share it to a Threat Intelligence Platform. This platform can allow users to locate known vulnerabilities in available AI models.

\vspace{-3mm}
\subsection{Generating AI Model Vulnerability Data}\label{fgm}

In \AlgName, we utilize Adversarial Robustness Toolbox (ART) \cite{gardtoolkits} available through DARPA's GARD, to generate AI model vulnerability data. ART utilizes a familiar setup process for Python programmers. Installation is performed over the Python package manager and facilitates installation of sub-options for popular AI/ML frameworks so that a user may install only what they need. 

ART requires a \textit{target victim ML model} for which the user wants to generate vulnerability data. ART includes many different wrapper-style classes for different neural network frameworks' implementations of models, including PyTorch and Tensorflow. This method of `importing' models allows compatibility with many different neural architecture implementations across frameworks. The wrappers allow for an implementation-agnostic utilization of the tools, alongside other class-wrappers for data loading - allowing training and testing of models.


In \AlgName, generating vulnerability data requires the user to provide a \textit{dataset} that would act as the source of perturbation for the attack models implemented in ART. ART comes with the ability to load certain popular neural network training datasets, but also implements its toolkit around many dataset-iterating classes.
ART abstracts much of the attack models that can be used, and utilizing it is as simple as passing the dataset and the model with some parameters specific to that attack method. For instance with ART's implementation of Fast Gradient Method (FGM), the toolkit provides parameter tuning options for the `epsilon' value, which is a scalar multiplier on the noise added to an image to create an adversarial example.


To demonstrate the generation of adversarial examples and vulnerability data using ART, we create FGM adversarial examples for the ResNet-50 convolutional network trained on CIFAR-10 data through the PyTorch framework. The same CIFAR-10 data was used as an input for ART. The first step of utilizing ART included leveraging its tools for loading the CIFAR-10 data into the existing data loading framework. The goal here was to perturb some of these input samples and cause model output miss-classification.    

\begin{figure}[h]
                \centering
                \includegraphics[scale=0.4]{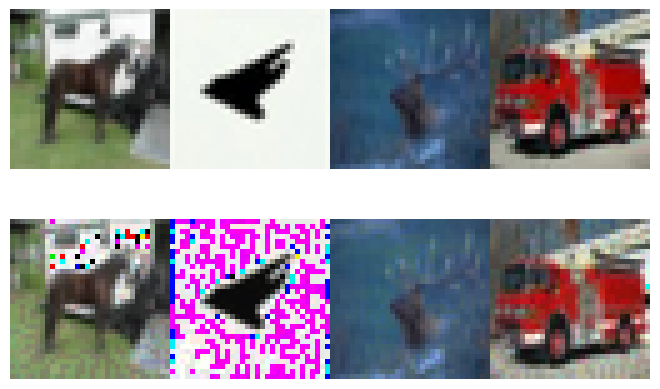}
                \caption{Comparison of normal (top) and FGM perturbed images (bottom) with $\epsilon$ set to 0.2.}
                \label{fig:perturbed_ep2}
                \vspace{-3mm}
            \end{figure}

With the loaded model and training CIFAR-10 set, we calibrated ART's FGM to the test data and generated adversarial examples. The model was then tested on the unaltered and altered images to find the accuracy difference imposed by FGM. The result of this process was a drop from 90.74\% accuracy to 44.41\% accuracy with an epsilon value of 0.2 (For some examples see Figure \ref{fig:perturbed_ep2}). 
The vulnerabilities identified in the model are then passed to the \AlgName's Threat Intelligence Encoder.

\subsection{Threat Intelligence Encoder (TIE)}\label{tie}

Once the vulnerability data has been identified, we pass the same to our Threat Intelligence Encoder (TIE). The encoder uses an adapted STIX format to create a shareable representation for the vulnerability data. We call the resulting format Artificial Intelligence Threat Information (AITI). The resulting threat intelligence can then be shared on different sharing platforms. Though for the \AlgName system, we utilize an adapted STIX format to encode the intelligence, it is also possible to encode the vulnerability data using variations of other industry standards like MISP, etc. 

To create AITI we first describe some \textit{domain objects} and their relations with each other. Some of the domain objects can be inherited in AITI from STIX \cite{stix1} like, Course of Action, Identity, Indicator, Observed Data, Threat Actor, Report, Vulnerability, etc. Additionally, in AITI, we include the following domain objects - 
\begin{itemize}
    \item \textit{AI Attack}: This object includes the information on the type of attack like, evasion, poisoning, model replication and exploiting traditional software flaws.
    \item \textit{AI Attack Pattern}: A type of Tactics, Techniques, and Procedures (TTPs) that describe ways that adversaries attempt to compromise target AI models. 

    \item \textit{Affected User Personas}: This information includes the type of users that can be affected by the particular AI threat. Potential values can be average user, security researchers, AI/ML researchers, etc.
    \item \textit{AI Paradigms under Threat}: This information contains details about how the target AI models have been developed and are operated. For example, AI models hosted on cloud, hosted on public servers, edge AI/ML models.
    \item \textit{AI Use Cases}: Information about AI systems and their uses. Include values like security-sensitive applications and non-security-sensitive applications. 
\end{itemize}

The other part of the AITI representation format is the relationship these domains have with each other. These relations are used to link together any two above mentioned domain objects. These links are referred to as \textit{relationship objects} \cite{stix1}. We first inherit the existing relations between STIX domain objects and include the following in AITI -

\begin{itemize}
    \item \textit{Relationship}: Used to link together two domain objects in order to describe how they are related to each other.
    \item \textit{Sighting}: Denotes the belief that something in the intelligence (e.g., an attack, attack patterns, tool, threat actor, etc.) was observed.

\end{itemize}

Using AITI, we can encode the vulnerability data generated in Section \ref{fgm} for the target victim model Resnet-50 trained on the CIFAR-10 dataset. The encoded example is available below:  

\begin{lstlisting}[language=json,firstnumber=1,caption = AITI object for vulnerability identified in Section \ref{fgm}]
{"type": "AI Attack-Evasion",
"id": "exampleFGM_Resnet-50_CIFAR10",
"created": "2022-08-11T23:39:03",
"AI Attack Pattern": "Fast Gradient Method (FGM) attack, hyperparameter: epsilon = 0.2",
"description": "An Fast Gradient Method (FGM) attack is possible against an object recognition AI model trained using the CIFAR-10 dataset based on the Resnet-50 architecture.",
"sophistication": "easy",
"resource_level": "individual",
"primary_motivation": "personal-gain"}
\end{lstlisting}

\vspace{-3mm}
\subsection{Threat Intelligence Sharing}

Once the vulnerability data has been encoded, it is now available to share among different shareholders. Development of a sharing platform directly compliments the need to enable communication between AI/ML model developers, users, and AI/ML security professionals. We propose adapting a RESTful API server like TAXII \cite{taxii} or the MISP threat sharing framework \cite{misp}. We believe augmenting existing tools and infrastructure that disseminates threat and vulnerability intelligence to include AI/ML vulnerability data is beneficial to for the overall ecosystem. Our AITI representation described in Section \ref{tie} is compatible with the existing TAXII services and message exchanges. 

\section{Conclusion \& Future Work}

In this paper, we have briefly described \AlgName, a work in progress system to identify and share AI/ML specific vulnerabilities and threat intelligence. The system leverages DARPA's GARD AI red teaming toolkit \cite{darpaFound} to identify vulnerabilities in an AI model. These vulnerabilities and threat intelligence are encoded in an adapted STIX object, called Artificial Intelligence Threat Information (AITI), that can then be shared using a RESTful API server like TAXII \cite{taxii}. Future works include expanding compatibility with other adversarial machine learning toolkits. AITI representation format needs to be further developed to handle complex threat intelligence representation. 

\section*{Acknowledgement}
Supported by National Science Foundation grant (\#2133190).

\bibliographystyle{unsrt}
\bibliography{refs}

\begin{thebibliography}{10}

\bibitem{darpaFound}
{DARPA} {Guaranteeing AI Robustness Against Deception (GARD)}.
\newblock
  \url{https://www.darpa.mil/program/guaranteeing-ai-robustness-against-deception}.

\bibitem{gardtoolkits}
{Adversarial Robustness Toolbox} ({ART}).
\newblock
  \url{https://adversarial-robustness-toolbox.readthedocs.io/en/latest/}.

\bibitem{trusted-ai}
Trusted-AI.
\newblock {ART attacks · trusted-ai/adversarial-robustness-toolbox wiki}.
\newblock
  \url{https://github.com/Trusted-AI/adversarial-robustness-toolbox/wiki/ART-Attacks}.

\bibitem{taxii}
{Trusted Automated Exchange of Intelligence Information (TAXII)}.
\newblock
  \url{https://oasis-open.github.io/cti-documentation/taxii/intro.html}.

\bibitem{stix1}
Oasis group.
\newblock {Structured Threat Information Expression} ({STIX}).
\newblock https://oasis-open.github.io/cti\-documentation/stix/intro.html, May
  2022.

\bibitem{misp}
MISP group.
\newblock {MISP} {Threat Sharing}.
\newblock https://www.misp-project.org, May 2022.

\bibitem{nvd}
{NIST}.
\newblock National vulnerability database.
\newblock https://nvd.nist.gov, May 2022.

\bibitem{mitreatlas}
{MITRE ATLAS, Adversarial Threat Landscape for Artificial-Intelligence
  Systems}.
\newblock \url{https://atlas.mitre.org}.
\newblock Accessed: 2022-08-01.

\bibitem{mitreattack}
{MITRE ATT\&CK}.
\newblock \url{https://attack.mitre.org}.
\newblock Accessed: 2022-08-01.

\bibitem{he2016deep}
Kaiming He, Xiangyu Zhang, Shaoqing Ren, and Jian Sun.
\newblock Deep residual learning for image recognition.
\newblock In {\em Proceedings of the IEEE conference on computer vision and
  pattern recognition}, pages 770--778, 2016.

\bibitem{krizhevsky2009learning}
Alex Krizhevsky, Geoffrey Hinton, et~al.
\newblock Learning multiple layers of features from tiny images.
\newblock 2009.

\bibitem{goodfellow_shlens_szegedy_2015}
Ian~J. Goodfellow, Jonathon Shlens, and Christian Szegedy.
\newblock Explaining and harnessing adversarial examples, Mar 2015.

\end{thebibliography}

\end{document}